\documentclass[twocolumn,english,superscriptaddress]{revtex4-1}
\usepackage{color,graphicx,natbib,subfigure,amsmath,amssymb,tabularx,epsfig,longtable,amsfonts,rotating,subfigure,hyperref,babel,currfile}
\begin{document}
\title{A Strategy for Preparing Quantum Squeezed States Using Reinforcement Learning}
\author{X. L. Zhao{$^*$}, Y. M. Zhao\footnote{These authors contributed equally to this work and should be considered co-first authors.}, M. Li, T. T. Li, Q. Liu, S. Guo,}
\affiliation{School of Science, Qingdao University of Technology, Qingdao 0532, China}
\author{X. X. Yi\footnote{Corresponding author\quad E-mail:~\textsf{yixx@nenu.edu.cn}}}
\affiliation{Center for Quantum Sciences and School of Physics,
Northeast Normal University, Changchun 130024, China}
\date{\today}
\begin{abstract}
We propose a scheme leveraging reinforcement learning to engineer
control fields for generating non-classical states. It is exemplified
by the application to prepare spin-squeezed states for an open
collective spin model where a linear control field is designed to
govern the dynamics. The reinforcement learning agent determines the
temporal sequence of control pulses, commencing from a coherent spin
state in an environment characterized by dissipation and dephasing.
Compared to the constant control scenario, this approach provides
various control sequences maintaining collective spin squeezing and
entanglement. It is observed that denser application of the control
pulses enhances the performance of the outcomes. However, there is
a minor enhancement in the performance by adding control actions.
The proposed strategy demonstrates increased effectiveness for
larger systems. Thermal excitations of the reservoir are detrimental
to the control outcomes. Feasible experiments are suggested to
implement this control proposal based on the comparison with the
others. The extensions to continuous control problems and another
quantum system are discussed. The replaceability of the reinforcement
learning module is also emphasized. This research paves the way for
its application in manipulating other quantum systems.
\end{abstract}
\maketitle

\section{Introduction}
Precise measurement for physical quantities propels the progress of physics.
The utilization of non-classical quantum states facilitates the pathway
towards achieving precise measurements. Spin-squeezed states are such
non-classical candidates characterized by reduced uncertainty in a collective
spin component. Entanglement typically arises concomitantly with spin squeezing,
as a result of the nonlinear interactions within an
ensemble~\cite{PR50989,RMP90035005}. Reducing the variance of an observable
enhances measurement sensitivity, surpassing the standard quantum limit in the
domain of quantum-enhanced metrology. Such squeezed states can be used to
enhance the performance of homodyne interferometers~\cite{PRA46R6797,PRA5067}, magnetometers~\cite{PRL109253605}, and atomic clocks~\cite{RMP87637}.
The evidence delineating the association between the sensitivity of
phase estimation and entanglement was demonstrated by a collective
spin model~\cite{PRL102100401}.

Many schemes have been proposed to prepare spin-squeezed states~\cite{PR50989,RMP90035005,PRA475138}. For example, quantum non-demolition measurement can be used to generate spin-squeezed states~\cite{EL42481}.
Theoretically, a coherent control method
has been proposed to produce spin-squeezed states~\cite{PRA63055601}.
Some proposals are based on the nonlinear interaction among the
individual elements in Bose-Einstein condensates~\cite{PRL853991,Nature40963,Nature4641165,Nature4641170}.
Proposals which can prepare long-lasting and extreme spin-squeezed
states are pursued all the way.

Machine-learning techniques are emerging as effective tools in
physics~\cite{Murphy2012,Sutton2018,Nature549195}, and among them,
reinforcement learning (RL) offers the potential to optimize the
control field for high-dimensional, multistage processes in complex
scenarios. Deep reinforcement learning (DRL) can provide a control
strategy to engineer the dynamics as long as the evolution follows
certain differential equations. In physics, many optimal problems
can be treated as control problems of finding means to steer a
system to achieve a certain target. The search for optimal control
fields can be formulated as RL tasks
\cite{Sutton2018,MURPHY2018,PRX8031084}. A semiquantum reinforcement
learning approach is employed to adapt a qubit to an unknown state
by successive single-shot measurements and feedback~\cite{AQT21800074}.
A four-step strategy combining constant-value and time-varying controls
results in an enhanced longtime stable spin squeezing for a collective
spin system coupled to a bosonic field~\cite{PRA103032601}. A
measurement-based adaptation protocol is introduced
to optimize a quantum state for maximum overlap with an unknown
reference state using quantum reinforcement learning~\cite{PRA98042315}.
This study provides a foundational framework for the development of
quantum reinforcement learning protocols leveraging quantum data.

We mainly propose an RL-based control scheme to design the control
field to prepare non-classical spin states. The agent is trained to
produce a sequence of square pulses that steer the system to squeezed
states under the domain of a Lindblad master equation. This control
method can be
regarded as a kind of combination of bang-bang
control~\cite{PRA582733,PRA86022321} and reinforcement learning. The
performance of the proposed control scheme is evaluated across a range
of control parameters, system size, and thermal excitations.

This paper is organized as follows: In Sec.\ref{ConStr}, we delineate
the reinforcement-learning-based framework to prepare  nonclassical
states. In Sec.\ref{RLandGA}, the RL module employed within the
control scheme is shown. In Sec.\ref{CSM}, we explicate the quantum
model for the generation of spin-squeezed states. In Sec.\ref{PSSS},
we present the procedure to prepare squeezed states of the open
collective spin system via reinforcement learning. In Sec.\ref{ConRes},
we check the performance of this method including the influence of
the frequency of applying the control pulses, the granularity of
control actions, the scalability across various sizes, the impact of
thermal excitation, and the comparison with other current protocols.
In Sec.\ref{discussion}, we discuss the feasibility of this control
scheme in terms of BEC experiments, the application to continuous
control situations and the extension to another quantum system. The
replaceability of the RL agent is also discussed. Finally, we conclude
in Sec.\ref{CONC}.

\section{Control Scheme}\label{ConStr}
Taking a cue from Lyapunov control strategies~\cite{24B.B42nd,25Auto411987,26AC50768,JPB44195503,PLA425127874},
we propose a scheme harnessing an RL agent to design the control
fields to prepare nonclassical states. In the presence of control
fields, the general total Hamiltonian for the quantum system reads:
$\hat{H}=\hat{H}_0+\sum_{m=1}^{M}f_m(t)\hat{H}_m$, where
$\hat{H}_0$ is the free Hamiltonian and $M$ denotes the amount of
the external control Hamiltonians $\hat{H}_m$. $f_m(t)$ is the
control field designed by the RL agent. It should be confirmed that
$[\hat{H}_{0},\hat{H}_{m}]\neq0$, otherwise, the influence of the
control Hamiltonians can be subsumed into the free Hamiltonian.

\begin{figure}
\includegraphics*[width=8.5cm]{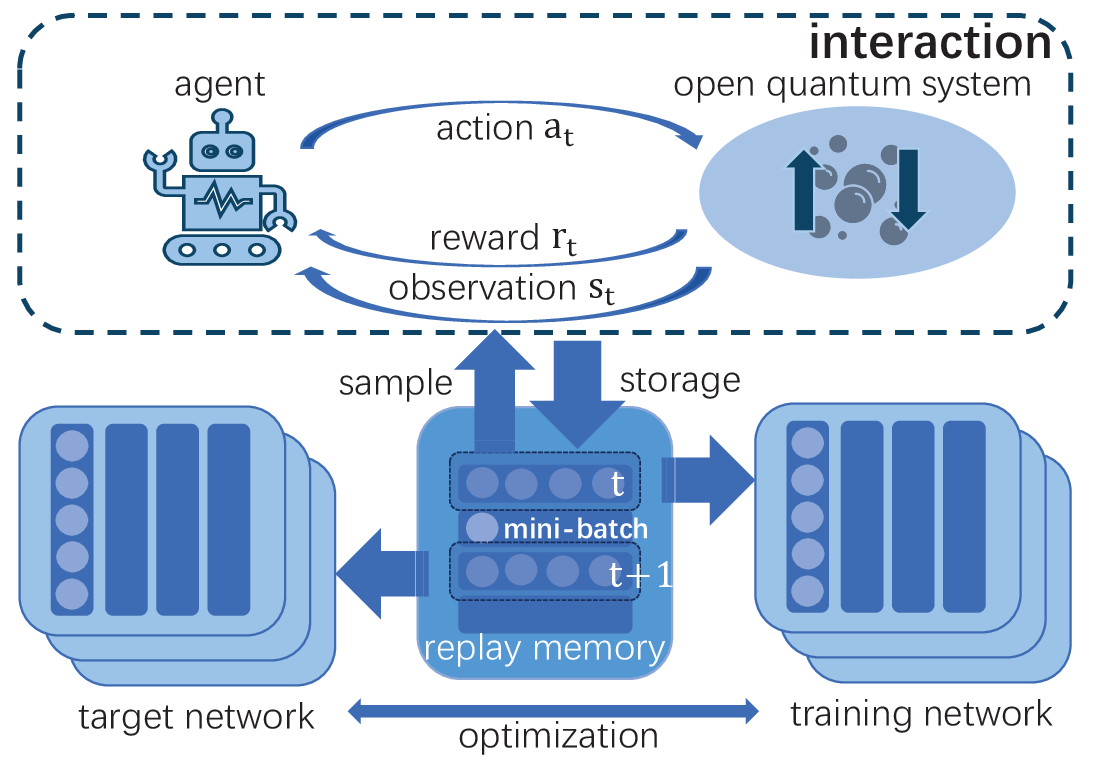}
\caption{The schematic of the training procedure: 1.
The agent provides actions $a_t$ to steer the open quantum system
(the `environment' in RL), and receives observations $s_t$ and rewards $r_t$.
2. These experiences are stored in the replay memory as a state-action-reward
combination of the mapping. 3. The agent is trained
to minimize the difference between the predicted Q values and the target
Q values. 4. The target Q values are computed by considering the immediate
reward and the estimated maximum future rewards. 5. The process including
interaction, experience replay, and network updates is repeated to iteratively
improve the Q-function approximation.}
\label{RLPro}
\end{figure}

This is an open-loop strategy in application, wherein the control
fields are determined through the emulation of a closed-loop
process, as shown in Fig.~\ref{RLPro}. Under the
control designed by the machine learning agent, the system would
be steered to a set of states meeting the control target. Notably,
the RL agent can be supplanted by other optimization modules
tailored to the target. The proposed control scheme can be applied
to other dynamical systems governed by certain differential
equations. To illustrate the efficacy of the scheme, we have
applied it to engineer a collective spin system.

\section{Machine Learning Tools: Reinforcement Learning}
\label{RLandGA}

Initiating with no prior knowledge about the system under control,
RL adopts the trial-and-error paradigm to iteratively learn the
mapping between the action and the state with maximal accumulated
reward over time instances. Rewards are obtained using an evaluation
rule that aligns with the control objectives. An appropriate
rewards-evaluation rule that favors particular state-action mappings
can enhance the control performance. Decision-making executed by
the agent entails selecting actions $a_t=\pi(s_t)$ that affect the
system changing from a state (not specifically referring to
quantum states, but a general state characterized by various
quantities of a controlled system) $s_t$ to $s_{t+1}$, with $\pi$
denoting the policy being learned~\cite{Nature549195,Murphy2012,Sutton2018}.

$Q$-learning works through a $Q$ function which
represents the expected total future reward of a policy $\pi$.
The optimal policy $\pi^*$ with maximized $Q$ function satisfies
Bellman equation~\cite{Nature518529,PNAS38716} which encapsulates
the principle of optimality for decision-making over time. Deep
$Q$-learning uses a deep neural network to approximate this function~\cite{Nature521436,GoodfellowDL,arXiv171002298},
and the network is called a deep Q network (DQN).

In DQN, the Q function is defined as
\begin{equation}
Q^\pi(s_t,a_t)=r(s_t)+E_{\{(s_{t+k},a_{t+k})|\pi\}_{k=1}^{\infty}}\!\left[\sum_{k=1}^{\infty}\gamma^i_{q}\, r(s_{t+k})\right]\label{Qp}
\end{equation}
where $r(s_t)$ is the reward at episode $t$ when the state is $s_t$, $\gamma^k_{q}$ ($0\!<\!\gamma^k_{q}\!<\!1$, usually close to 1) is
a discount factor, and the expectation $E_{\{\cdot\cdot\cdot\}}[\cdot\cdot\cdot]$
is taken over trajectories of the actions and the states of the controlled system~\cite{Nature518529,PNAS38716}. The Q function is implemented during
training based on the structure of the agent. When the Bellman equation
\begin{equation}
Q^{\pi^*}(s_t,a_t)=r(s_t)+\gamma_{q}\,E_{s_{t+1}}
\left[\max_{a_{t+1}}Q^{\pi^*}(s_{t+1},a_{t+1})\right]\label{Qps}
\end{equation}
for $\gamma^k_{q}=\gamma_q \approx 1$, is satisfied by training,
the optimal policy $\pi^*$ (state-action mapping) is determined.
Deep Q-learning employ a neural network, denoted by $W_{\theta}(s,a)$,
to approximate $Q^{\pi^*}$ \cite{Nature521436,GoodfellowDL}, where
$\theta$ represents the network parameters that need to be adjusted
by training. After training, $W_{\theta}(s,a)$ approximates
$Q^{\pi^*}$ by adjusting the parameters $\theta$ so that
Eq.~(\ref{Qps}) is satisfied. Then the optimal policy $\pi^*$
be found.

DQN leverages extensive datasets that encapsulate the anticipated
cumulative rewards consequent to specific actions within distinct
states, facilitating the neural network-based approximation of
the conventional action-value function \cite{Nature521436,GoodfellowDL,arXiv171002298,CS2500611,PMLR387395,
PMLR19281937,arXiv151106581}.
It should be noted that depending on particular needs and
constraints of tasks, alternative optimization algorithms
could be employed in place of the DQN.

\section{Collective Spin Model}
\label{CSM}
We consider an ensemble of $N$ identical two-level
atoms with pseudo spin components $\hat{J}_{\alpha} = \frac{1}{2}\sum^{N}_{k = 1}\hat{\sigma}_{\alpha}^{(k)}$,
$(\alpha = x, y, z)$, where $\hat{\sigma}_{\alpha}^{(k)}$ is the Pauli
operator for the $k$-th atom~\cite{PRA62211}. This is the symmetric scenario
where the operations done on the ensemble have identical
impact on all the atoms.
$\hat{J}_x$, $\hat{J}_y$, $\hat{J}_z$ fulfill the SU(2) commute relationship:
$[\hat{J}_{\alpha},\hat{J}_{\beta}] = i\hbar\epsilon_{\alpha\beta\gamma}\hat{J}_{\gamma}$,
where $\epsilon_{\alpha\beta\gamma}$ is the L\'evi-Civit\`a symbol. The total
collective spin length is specified by $J = N/2$ and the dimension
of the Hilbert space is $2J+1 = N+1$. The collective spin can be mapped to
its two-mode bosonic partner by Schwinger transformation:
$\hat{J}_z = \frac{1}{2}(\hat{a}^\dagger\hat{a}-\hat{b}^\dagger\hat{b})$,
$\hat{J}_+ = \hat{a}^\dagger\hat{b}$ and $\hat{J}_- = (\hat{J}_+)^\dagger$,
where $\hat{a}$ and $\hat{b}$ are the two annihilation operators of two
boson modes. Namely, $\hat{J}_x = (\hat{J}_++\hat{J}_-)/2$, and
$\hat{J}_y = (\hat{J}_+-\hat{J}_-)/2i$~\cite{AMQP1981}. In
this view, by mapping one mode to spin up and the other one to spin down,
$\hat{J}_z$ reflects the population difference between the two modes in
Ramsey interferometer~\cite{PRA46R6797,PRA475138,RMP90035005}.

To examine the efficacy of the strategy, we focus on a collective spin
system described by the Hamiltonian
\begin{equation}
 \hat{H}/\hbar = \kappa \hat{J}_{z}^{2}+\Omega_x(t)\hat{J}_{x},
 \label{HC}
\end{equation}
here $\kappa$ indicates the strength of the interaction between the
atoms and the time scales as $\kappa t$. $\kappa$ would be taken as
the unit ($\kappa = 1$ hereafter) and the natural units ($\hbar = 1$)
is used. $\hat{J}_z^2$ refers to the one-axis twisting which induces
spin squeezing~\cite{PRA475138}. This squeezing provides the resource
for quantum-enhanced metrology~\cite{PRA475138,RMP90035005}. $\hat{J}_{x}$
is the control Hamiltonian describing the magnetic field in the $x$-direction,
or the counterpart, the linear beam splitter in interferometers.

In contrast to the application of a constant control~\cite{PRA63055601},
the present proposal employs an RL agent designing
time-dependent control field $\Omega_x(t)$ to prepare nonclassical states.
The temporally varying control field can represent the operational analogy
of linear beam splitters in interferometric experiments~\cite{Nature4641165,Nature4641170}. Since
$[\hat{J}_z^2,\hat{J}_{x}] = i(\hat{J}_{y}\hat{J_z}+\hat{J}_z\hat{J}_{y})$,
such a linear control Hamiltonian can be used to steer the spin system.

Spin squeezing can be quantified by parameters constructed by the expected
values of collective spin operators~\cite{PR50989,RMP90035005}.
Upon the reduction of the variances beneath the standard quantum limit
threshold, the system is rendered applicable for precision metrology,
along with amplifying the variance of the orthogonal spin components.
We should confirm that the minimum squeezing parameter reads
\begin{equation}
\begin{split}
\xi_{\perp}^{2}& = \frac{N \min(\Delta\hat{J}_{\vec{n}_{\perp}}^{2})}{|\langle \hat{J}_s\rangle|^2}\\ & = \frac{N\left[\left\langle \hat{J}_{\vec{n}_{1}}^{2}+\hat{J}_{\vec{n}_{2}}^{2}\right\rangle -\sqrt{\left\langle \hat{J}_{\vec{n}_{1}}^{2}-\hat{J}_{\vec{n}_{2}}^{2}\right\rangle ^{2}+\langle \left[\hat{J}_{\vec{n}_{1}},\hat{J}_{\vec{n}_{2}}\right]_+\rangle^{2}}\right]}{2|\langle \hat{J}_s\rangle|^2},\label{XR}
\end{split}
\end{equation}
where $\hat{J}_{\vec{n}_{i}}$ = $(\hat{J}_x,\hat{J}_y,\hat{J}_z)\cdot \vec{n}_{i}$,
($i = 1,2$) and $\vec{n}_{1} = (-\sin{\phi},\cos{\phi},0)$,
$\vec{n}_{2} = (\cos{\theta}\cos{\phi},\cos{\theta}\sin{\phi},-\sin{\theta})$. The collective spin operator $\hat{J}_s$ = $(\hat{J}_x,\hat{J}_y,\hat{J}_z)\cdot (\sin \theta \cos\phi,\sin \theta \sin \phi, \cos \theta)$, where $\theta$ = $\arccos(\frac{\langle
\hat{J}_{z}\rangle}{|{\hat{J}}|})$,
$\phi$ = $\mathrm{sign}(\langle\hat{J}_y\rangle)\arccos(\frac{\langle \hat{J}_{x}\rangle}{|{\hat{J}}|\sin\theta})$, the norm $|{\hat{J}}| = \sqrt{\langle \hat{J}_{x}\rangle^{2}+\langle \hat{J}_{y}\rangle^{2}+\langle \hat{J}_{z}\rangle^{2}}$ \cite{PR50989,RMP90035005,JPB39559}. Here the direction $\vec{n}_{\perp}$ = $\vec{n}_{1}\cos\varphi+\vec{n}_{2}\sin\varphi$ with
\begin{equation}
	\varphi=\left\{ \begin{array}{ll}
		\frac{1}{2}\arccos\Big(\frac{-A}{\sqrt{A^{2}+B^{2}}}\Big) & \text{if}~~B\leq0,\\
		\pi-\frac{1}{2}\arccos\Big(\frac{-A}{\sqrt{A^{2}+B^{2}}}\Big) & \text{if}~~B>0,\end{array}\right.\label{minimumang}
\end{equation}
where $A\equiv\langle \hat{J}_{\vec{n}_{1}}^{2}-\hat{J}_{\vec{n}_{2}}^{2}\rangle,\text{ \ }\; B\equiv\langle \left[\hat{J}_{\vec{n}_{1}},\hat{J}_{\vec{n}_{2}}\right]_+\rangle$.

In this control scheme, we employ the following definition
\begin{equation}
\xi_{Z}^2 = \frac{N\Delta \hat{J}_{z}^2}{|\langle \hat{J}_s\rangle|^2},
\label{Xis}
\end{equation}
as the squeezing parameter. Here $\Delta \hat{J}_{z}^2$ =
$\langle \hat{J}_z^2\rangle-\langle \hat{J}_z\rangle^2$, indicates the spin
squeezing in the z-direction. For the RL, we use the reverse of $\xi_{Z}^2$
to set the reward-punishment rule during the control process. The reward
function we use is
$R = 10\langle\Delta R_t \geq 0\rangle-\langle\Delta R_t < 0\rangle $, where
\begin{equation}
	\langle\bullet\rangle = \left\{
\begin{array}{ll}
		~~~0& \text{if}~~\bullet False ,\\
		~~~1& \text{if}~~\bullet Truth,
\end{array}\right. \label{sign}
\end{equation}
$\Delta R_t = \frac{1}{\xi_z^2}_{(t+1)}-\frac{1}{\xi_z^2}_{(t)}$. Here $\frac{1}{\xi_z^2}_{(t)}$ is the inverse of the squeezing parameter at the
sample time $t$ during training. This reward function means that the agent
gets 10 points if the action makes the squeezing parameter decrease, whereas
deduct 1 point if the squeezing parameter rises. Subsequently, in accordance
with the fundamental principles of RL, the agent will steer the direction
with minimum squeezing parameter approaching the $z$ direction, namely,
$\varphi = \pi/2$ in Eq.~(\ref{minimumang}). Compared
to using the reverse of $\xi_{R}^{2}$ with variable $\theta$ and $\phi$,
the choice of $\xi_{Z}^2$ possesses a more direct physical interpretation
since $\hat{J}_z$ signifies the population imbalance between the two modes
within the Ramsey interferometer mentioned above~\cite{PRA46R6797,PRA475138,RMP90035005}.

Correlation exists between entanglement and spin squeezing~\cite{PRL864431},
wherein multipartite entanglement constitutes a quantum resource for
enhanced precision in metrology~\cite{PRL102100401,PRA85022322}.
Moreover, quantum Fisher information (QFI) quantifies the link between
entanglement and phase uncertainty within the domain of metrology ~\cite{JSP1231,Science345424}. Adhering to the quantum Cramer-Rao
bound, quantum states with larger QFI are pursued for the precision of
quantum metrology~\cite{PLA25101,PRL723439}. We would check the
QFI about the spin state $\hat{\rho}$ concerning
$\hat{\rho}(\delta) = e^{i\delta\hat{G}}\hat{\rho}e^{-i\delta\hat{G}}$,
where $\delta$ is the quantity which needs to be estimated with respect
to the phase-shift operator $\hat{G}$~\cite{PRA88043832,SR58565}.
The QFI reads
\begin{equation}
F(\rho,\hat{G}) = 4\sum_{n}p_{n}(\Delta \hat{G})_{n}^{2}-\sum_{m\neq n}\frac{8p_{m}p_{n}}{
p_{m}+p_{n}}|\langle \psi _{m}|\hat{G}|\psi _{n}\rangle |^{2}, \label{QFI}
\end{equation}
where $\rho|\psi_{n}\rangle$ = $p_{n}|\psi_{n}\rangle$,
$(\Delta \hat{G})_{n}^{2}\equiv\langle\psi_{n}|\hat{G}^{2}|\psi_{n}\rangle-|\langle\psi_{n}|\hat{G}|\psi_{n}\rangle|^{2}$.
The second term denotes a correction. Here, QFI provides a quantitative
threshold for the precision attainable in estimating $\delta$ by measuring
$\hat{G}$ on $\rho$. If the average QFI over three basic directions reaches
the order of 1, there is macroscopic multi-particle entanglement~\cite{PRA85022321}.
A comprehensive solution for QFI calculation has already been integrated
into the package QuanEstimation~\cite{PRR4043057}.

\section{Prepare Spin-squeezed State by Reinforcement Learning}
\label{PSSS}
There are mainly two steps to prepare the spin-squeezed states:
firstly, a spin coherent state should be prepared, and secondly,
a spin-squeezed state is prepared by using a control field
designed by a machine learning agent.
\subsection{Initial coherent spin state}
$N$ two-level atoms all pointing along the same
direction can be described by SU(2) coherent spin state (CSS).
Such a state reads
\begin{equation} |\theta,\phi\rangle = (\cos\frac{\theta}{2})^{2j}\sum_{m = -j}^{j}(C_{j+m}^{2j})^{1/2}
	 [e^{-i\phi}\tan\frac{\theta}{2}]^{j+m}|j,m\rangle,
\label{initialstate}
\end{equation}
where $C_{j+m}^{2j}$ are the binomial coefficients. This overcomplete
state is most similar to the classical one with $\theta$ and $\phi$
being the azimuth angles for longitude and latitude, respectively.
$|j,m\rangle$ are the eigenvectors that satisfy the equations
$\hat{J}^2|j,m\rangle$=$j(j+1)\hbar^2|j,m\rangle$ and
$\hat{J}_z|j,m\rangle$=$m\hbar|j,m\rangle$ ($\hbar$=1 in numerical
calculations). The quantum state can be represented by the Husimi
function or the Wigner distribution. The CSS can be prepared by
applying $\pi/2$ pulses to a BEC with $N$ atoms in the internal
ground state~\cite{Nature40963,Nature4641165,Nature4641170}.
In the CSS, $\langle \hat{J}_{x}\rangle = N/2$ and $\langle \hat{J}_{y}\rangle = \langle \hat{J}_{z}\rangle = 0$. Such a pulse
is equivalent to the effect of a beam splitter in an interferometer.

\begin{figure*}
	\includegraphics*[width=18cm]{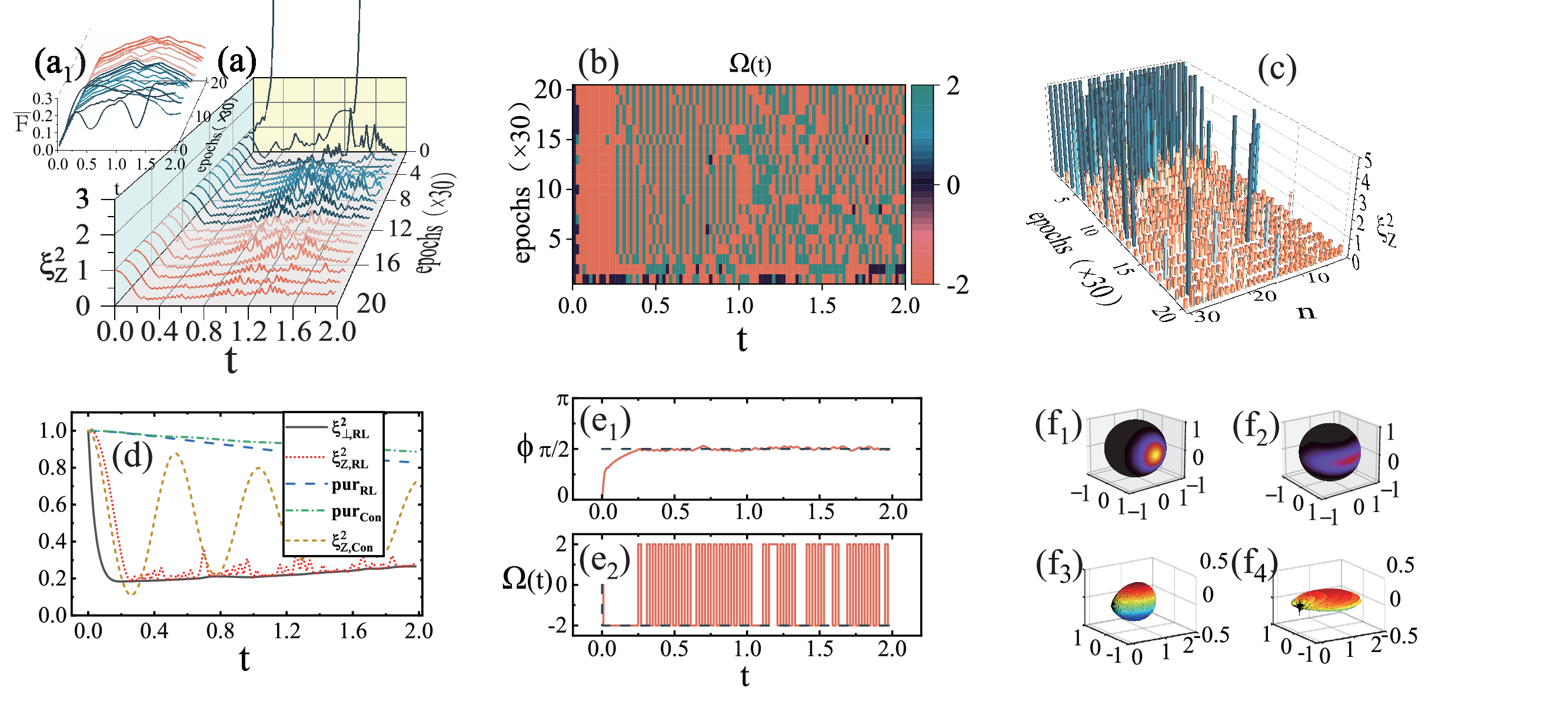}
	\caption{(a) Evolution of the spin squeezing parameter (\ref{Xis})
versus training epochs while 100 segments of actions are evenly taken
across each evolution time interval $[0,2]$. Every 30 training epochs
are shown within 600 consecutive training rounds. The subfigure in
(a$_1$) shows the evolution of the corresponding averaged QFI defined
as $\bar{F}$ = $[F(\rho,\hat{J}_x)+F(\rho,\hat{J}_y)+F(\rho,\hat{J}_z)]/(3N^2)$.
(b) Top view of the 3-action ($\Omega(t) = 2, 0, -2$) square control
pulse corresponding to the spin squeezing curves in (a). (c) 30 samples
(differentiated by $n$) of the squeezing parameter at $t=2$ of those
in (a). As shown in Fig. (a), there are control sequences causing
$\xi_{Z}^2$ to diverge at the beginning. However, such control sequences
are invalid. Therefore, we express instances where $\xi_{Z}^2 > 5$
uniformly as 5 to facilitate the visualization of the results in (c).
(d) The squeezing parameters versus time for the constant coherent
control ($\Omega(t) = -2$) and the RL-designed control, with the
corresponding degree of mixing indicated by $Tr[\rho^2(t)]$. The
subscript RL means $\Omega(t)$ obtained by reinforcement learning,
Con means $\Omega(t) = -2$. (e$_1$) evolution of the angle $\varphi$
in Eq.~(\ref{minimumang}). (e$_2$) The control fields $\Omega(t)$
designed by the RL agent corresponding to (d) and the constant control
field $\Omega(t) = -2$. (f$_1$)and(f$_2$) are the Husimi representation
of the initial CSS and the spin squeezed state at $t = 2$. (f$_3$) and
(f$_4$) are the polar plots of the Wigner functions for the states.
We use $N = 2J = 20$ hereafter. The decaying parameters are
$\gamma = \gamma_z = 0.001\kappa$.}
\label{DoubleCFxi}
\end{figure*}

Commencing with a CSS aligned along the $x$-axis and characterized by
isotropic fluctuation in its spin components, $\hat{J}_z^2$ shears the
coherent state to a squeezed one with the reduced variance of $\hat{J}_z$,
culminating in the generation of a spin-squeezed state. Such states
exceed the constraints delineated by the standard quantum limit,
allowing for enhanced measurement sensitivity in metrology along the
squeezed direction~\cite{PRL102100401}. The squeezed direction would
be fixed on $\hat{J}_z$ under the action of $\Omega(t)\hat{J}_x$
determined by the RL agent as mentioned above.

\subsection{Prepare spin-squeezed states by machine-designed pulses}

Usually, it is hard to avoid decoherence in a quantum system due
to its interaction with the environment. The effect of such decoherence
should be taken into account in the control scheme. We consider two
kinds of decoherence channels: superradiant damping and dephasing.
Such decoherence channels lead to the loss of quantum resources. The
time evolution of the collective spin system is described by the
Lindblad master equation as
\begin{eqnarray}
\dot{\rho}_{\rm} = -i[\hat{H}_{\rm},\rho_{\rm}] + \gamma(n_{th}+1) \mathcal{L}_{\hat{J}_-}\rho_{\rm}+ \gamma n_{th} \mathcal{L}_{\hat{J}_+}\rho_{\rm}
+  \gamma_z \mathcal{L}_{\hat{J}_z}\rho_{\rm} ,
\label{eqmaster}
\end{eqnarray}
where $\mathcal{L}_{\hat{X}}\rho = 2 \hat{X}^{\dag} \rho \hat{X} - \hat{X} \hat{X}^{\dag} \rho - \rho \hat{X} \hat{X}^{\dag}$. $\gamma$ is the
decay rate, $\gamma_z$ is the dephasing rate and $n_{th}$ is the
average thermal photons. Different from the traditional quantum Lyapunov
control strategies which are based on the distance between
eigenstates~\cite{PLA425127874}, the system would evolve under the
domain of this master equation with the application of the control
field $\Omega(t)$ designed using RL in the Hamiltonian (\ref{HC}).

In this work, the RL agent selects those actions contingent upon the
observations, thereby orchestrating a sequence of actions aimed
at maximizing cumulative rewards and minimizing penalties. During
the training, the observation (calculated based on the quantum state)
is fed to the neural network, while output neurons provide the
probability of choosing which action at each iterative training
step. A reward is dispensed subsequently to each step to evaluate
the decision-making policy. After one epoch, the collective spin
system is re-initialized to the coherent state and the next epoch
starts to train the agent continuously based on the trained neural
network. The state evolves deterministically according to the
master equation~(\ref{eqmaster}). The pulse strengths and
application time in the episode represent the policy $\pi^*$.

There are means to improve the performance of
an RL agent. The replay mechanism stores the learned history in
training, and enhances the learning efficiency and stability.
Besides, we employ Huber loss in the RL agent~\cite{AMS3573}
since it is robust when the error becomes too large due to the
linear function used. For a batch of $N$ samples, the Huber
loss is defined as: $L=\{l_1,...,l_N\}^T$, where
\begin{equation}
	l_n = \left\{ \begin{array}{ll}
		\frac{1}{2}e^2 & \text{if}~~|e|\leq\Delta,\\
		\delta (|e|-0.5*\Delta) & \text{if}~~|e|>\Delta,\end{array}\right.\label{Hubber}
\end{equation}
Here, $e$ represents the error term, for instance, the difference
between the predicted and actual Q values, $Q(s,a) - Q^*(s,a)$,
and $\Delta$ is a hyperparameter. It makes the training more
stable and provides a balance between Mean Squared Error
(underestimates large errors) and Mean Absolute Error
(overestimates small errors). It also helps in reducing the
exploding gradients problem in training deep neural networks
and can potentially lead to faster convergence compared to the
other loss functions~\cite{AMS3573}.

The neural network parameters are updated
 by the descent method called AdamW~\cite{60RfadamW}. The
 key difference between Adam and AdamW lies in their approach
 to weight decay which helps prevent over-fitting by adding
 a penalty term to the loss function. Huber loss and AdamW
 can be replaced by other loss functions and Optimization algorithms~\cite{Murphy2012,Sutton2018}.

As a result of this training, the weights of the neural network
are adjusted, i.e., the agent learns to determine a sequence
of actions based on the states of the system to obtain a larger
reward. Randomness provides the probability for the RL agent to
find the best sequence of actions.

It is imperative to clarify that the strategy is a closed-loop
simulation, but an open-loop application scheme. Once the control
sequence of actions, such as $\Omega_{x}(t)$ delineated in
Eq.(\ref{HC}), is obtained by simulation, the identical control
field is implemented in an open-loop control process to circumvent
quantum collapse attributable to the observation on the system.

\section{Control Results}\label{ConRes}
\subsection{The results of the control method}
As depicted in Fig.~\ref{DoubleCFxi}, the time interval [0,2] is
partitioned into a variable number of segments, and the control
square pulse is applied at the boundaries between the adjacent
time segments and sustained until reaching the subsequent boundary.
The RL agent designs the application of the pulses. Each round
of the training consists of 600 epochs. Every 30 epochs of the
evolution for the squeezing parameter are illustrated in
Fig.~\ref{DoubleCFxi} (a). It can be seen that,
at the early stages of the training, the agent does not provide an
effective control strategy, resulting in the divergence of $\xi_Z^2$.
However, as the training proceeds, the RL agent can find numerous
sequences of square pulses inducing a reduction of the squeezing
parameter. Meanwhile, the inset Fig.~\ref{DoubleCFxi} $(a_1)$
delineates the evolution of the corresponding averaged QFI. As
the training progresses, $\bar{F}$ tends to attain a high value
and descend gradually. This descent corresponds to the increase of
$\xi_Z^2$ resulting from the decoherence. Fig.~\ref{DoubleCFxi} (b)
depicts the corresponding square wave control pulses from the
top view. To show the efficacy of the control strategy from another
view, Fig.~\ref{DoubleCFxi} (c) reveals that, upon incrementing
the training epochs, there is a discernible decrease in the
squeezing parameter at $t = 2$ from the statistical view. These
results corroborate the efficacy of the control pulses designed
by the RL agent in optimizing the control performance.
\begin{figure}
	\includegraphics[width=8.6cm]{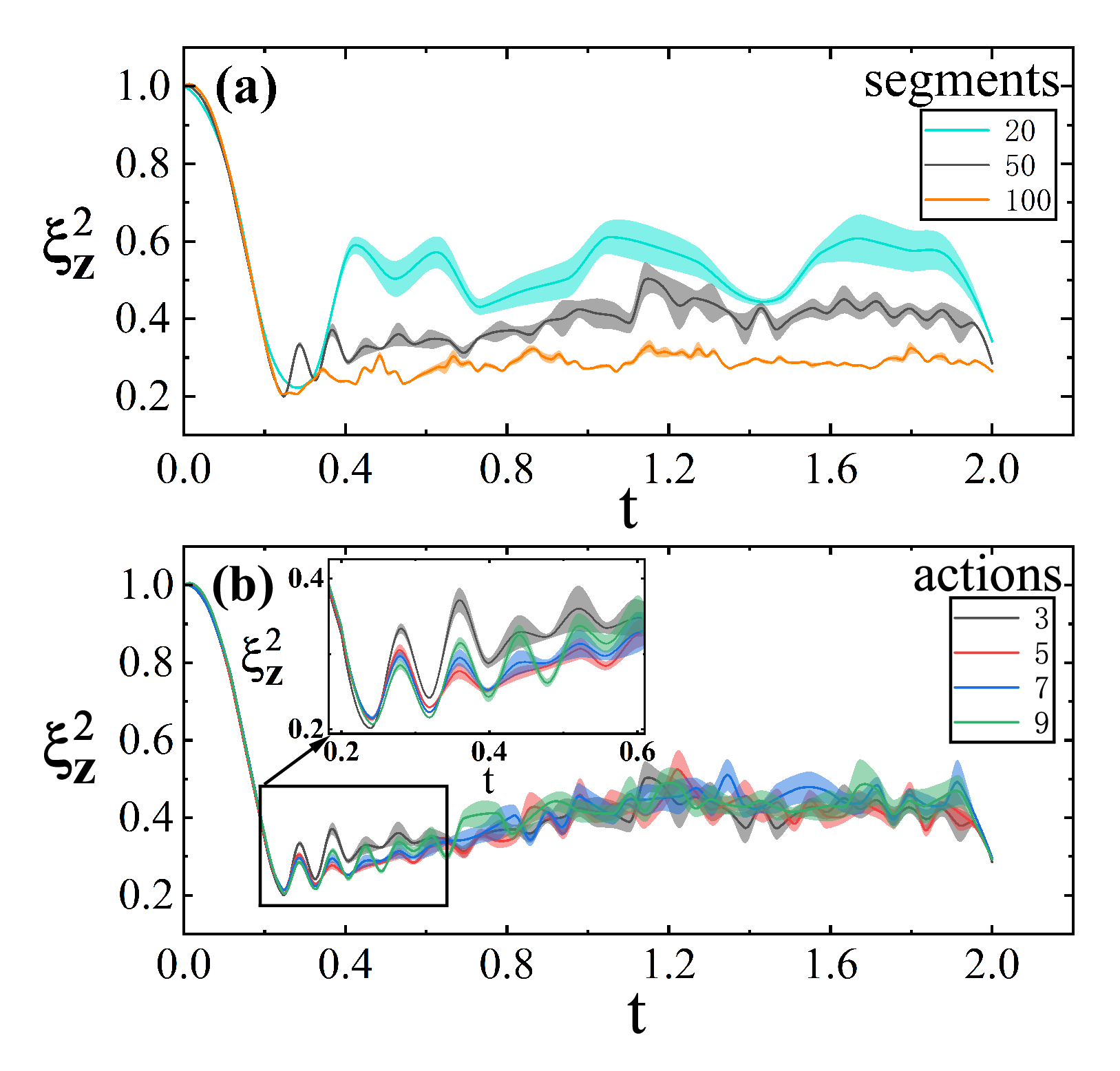}
	\caption{(a) Mean evolution of 30 samples of the spin squeezing
parameter $\xi_{Z}^2$. The shaded regions indicate the standard errors
of the fits for the different frequencies of applying the control
pulses. Control time interval $[0,2]$ is evenly divided into the
number of segments at which the square pulse sequence
with amplitude 2 ($\Omega(t) = 2,0,-2$) is applied. We should confirm
that each of the 30 samples is the one with the smallest $\xi_{Z}^2$
among the 600 epochs as in Fig.~\ref{DoubleCFxi}. It is reasonable
since in practical scenarios, one should choose the best control
sequence in experiments. This results in the contraction of the
fluctuation error values at time $t = 2$. (b) Evolution of the spin
squeezing parameter for different numbers of control actions when the
number of segments is 40:
$\{\Omega(t)\}$ =
$\{(2, 0, -2)|~actions = 3\}$,
$\{(2, 1, 0, -1, -2)|~actions = 5\}$,
$\{(2, 1.34, 0.66, 0, -0.66, -1.34, -2)|~actions = 7\}$,
$\{(2, 1.5, 1, 0.5, 0, -0.5, -1, -1.5, -2)|~actions = 9\}$.
The other parameters are the same as those in Fig.~\ref{DoubleCFxi} (a).}
\label{DoubleCFeacc}
\end{figure}
Fig.~\ref{DoubleCFxi} (d) presents the evolution of the spin squeezing
parameter obtained by using RL-designed $\Omega(t)$: $\xi_{Z,RL}^{2}$,
constant control field $\Omega(t) = -2$: $\xi_{Z,Con}^{2}$, and the
optimized $\xi_{\perp,RL}^2$, with the minimal value at the
final control time $t = 2$ among the 600 epochs. The comparison
clearly demonstrates a significantly enhanced performance of the
RL-based control approach as opposed to the constant control
scenario under the same parameter settings. We employ the trace
of the square of the system state $\rho(t)$: $Tr[\rho^2(t)]$ to
describe the degree of mixing for a quantum state. It can be seen
that the degree of mixing of the state under the RL control tends
to be more deeply compared to $\Omega(t) = -2$. The optimal squeezing
angle $\varphi\rightarrow \pi/2$, i.e., converges to the z-component,
which can be seen in Fig.~\ref{DoubleCFxi} $(e_1)$. The linear term $\Omega_x(t)\hat{J}_{x}$ rotates the fluctuation and $\hat{J}_z^2$
twist the fluctuation. The combination of these two terms leads to
long-lasting spin-squeezed states. The square wave control
sequence corresponding to the RL control result in
Fig.~\ref{DoubleCFxi} (d) is depicted in Fig.~\ref{DoubleCFxi} $(e_2)$.
Furthermore, Fig.~\ref{DoubleCFxi} $(f_1)-(f_4)$ utilize the Husimi
function and the Wigner function to visualize the initial coherent
spin state and final spin-squeezed state at the time $t=2$.
Nonclassical states are characterized by the twisted distribution
in Husimi function~\cite{PR50989,PRA475138} and the asymmetry
in the polar plot of Wigner function~\cite{PRA604034,PRL117180401}.
This provides insight into the evolution of the quantum state
under control. To show the squeezing process more
vividly, a movie of the squeezing process is shown by Husimi
function in~\cite{Movie}.

We conjecture that the application frequency of the pulses impacts
the outcomes of the control. To investigate the evolution of the
squeezing parameter versus different application frequencies,
we split the temporal interval [0, 2] into different numbers
of segments, the more segments, the more frequently applying the
pulses. It can be seen in Fig.~\ref{DoubleCFeacc}(a), with increasing
the segments, the squeezing parameter is depressed more stationary
and lower. Even more, with increasing of the control pulses, the
variance of the squeezing parameter is also depressed more obviously.
Straightforwardly, there is a contradiction between the control
performance and operation difficulty. One may conjecture the number
of control actions also influences the control performance. However,
as shown in Fig.~\ref{DoubleCFeacc}(b), there is no obvious advantage
for more control actions with the same maximum control amplitude in
this control.

\begin{figure}
	\includegraphics*[width=8.6cm]{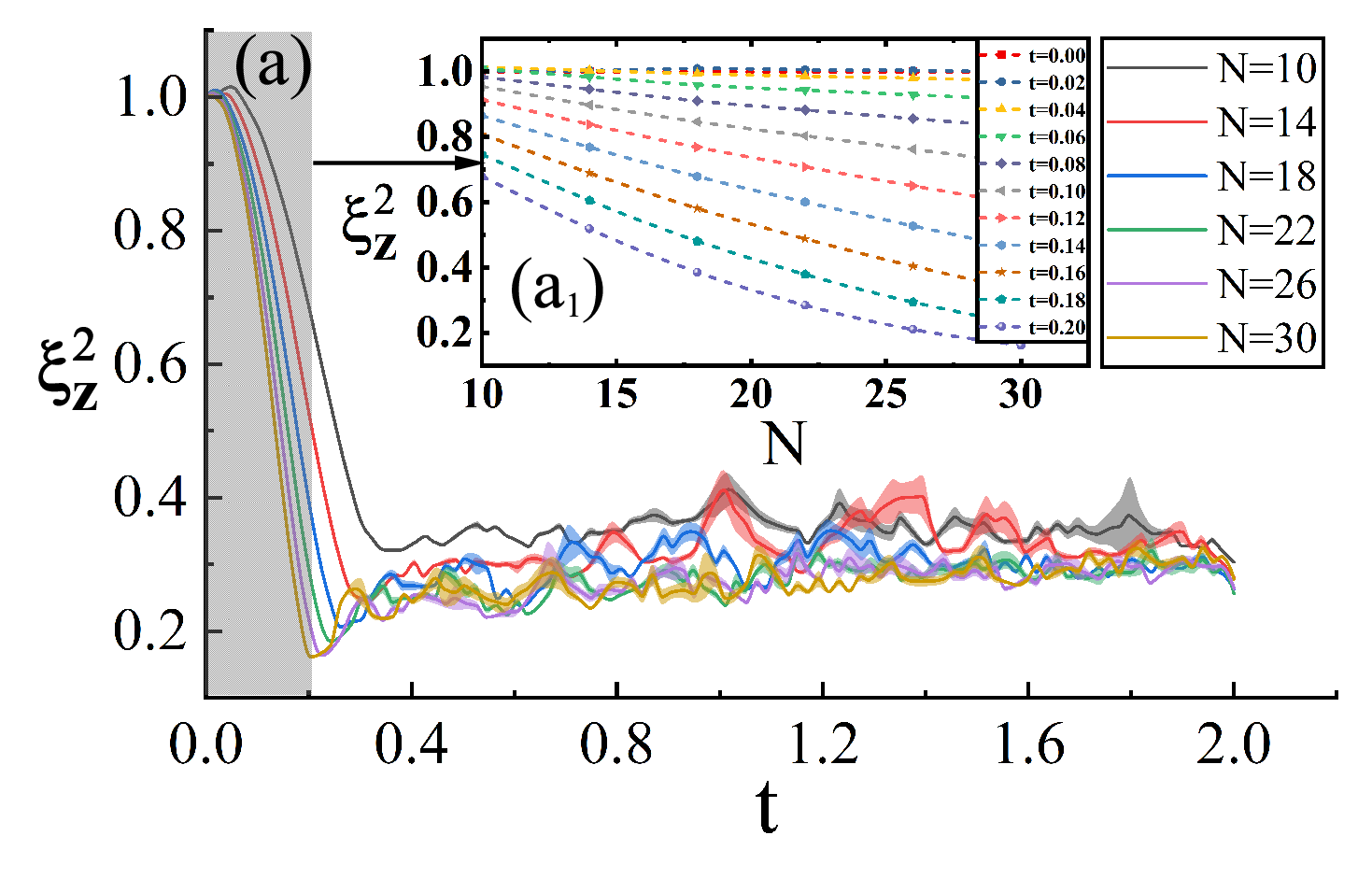}
	\caption{Average evolutions for 30 samples of the spin squeezing
parameter $\xi_{Z}^2$ with error zones for different sizes of the
collective spin system $N = 2J$ under 3-action ($\Omega(t)=2,0,-2$)
control. The samples are picked in the same manner as those shown
in Fig.~\ref{DoubleCFeacc} (a). The subgraph shows the squeezing
parameter in the scaled time interval $t = [0,0.2]$, picked from the
shaded area on the left in the graph, for the collective spin system
of different sizes. The same color is shared with the same $N$. The other
parameters are same to those in Fig.~\ref{DoubleCFxi} (a).}
\label{RLlearning3size}
\end{figure}

It is natural to ask about the applicability of the proposed scheme
to collective spin models with different total spin numbers ($N = 2J$).
To address this concern, Fig.~\ref{RLlearning3size} ($a$) illustrates
the control result for the collective spin systems with different $N$s.
The result reveals that the squeezing parameter reaches lower values
for larger $N$ with the same other parameters. This observation
suggests that an enlarged ensemble of spins benefits enhancing the
precision of quantum metrology. Furthermore, an examination of the
subgraph discloses a convergence towards parallelism among the
trajectories corresponding to different $N$s, indicating an emergence
of scaling behavior.
In the system under consideration, energy dissipation concurrent
with decoherence interplays with the applied coherent pulses which
impedes the quantum system decay to the ground state. Consequently,
the plateau in the squeezing parameter can be attributed to the
dynamic equilibrium between these two conflicting factors.

\begin{figure}
\includegraphics*[width=8.6cm]{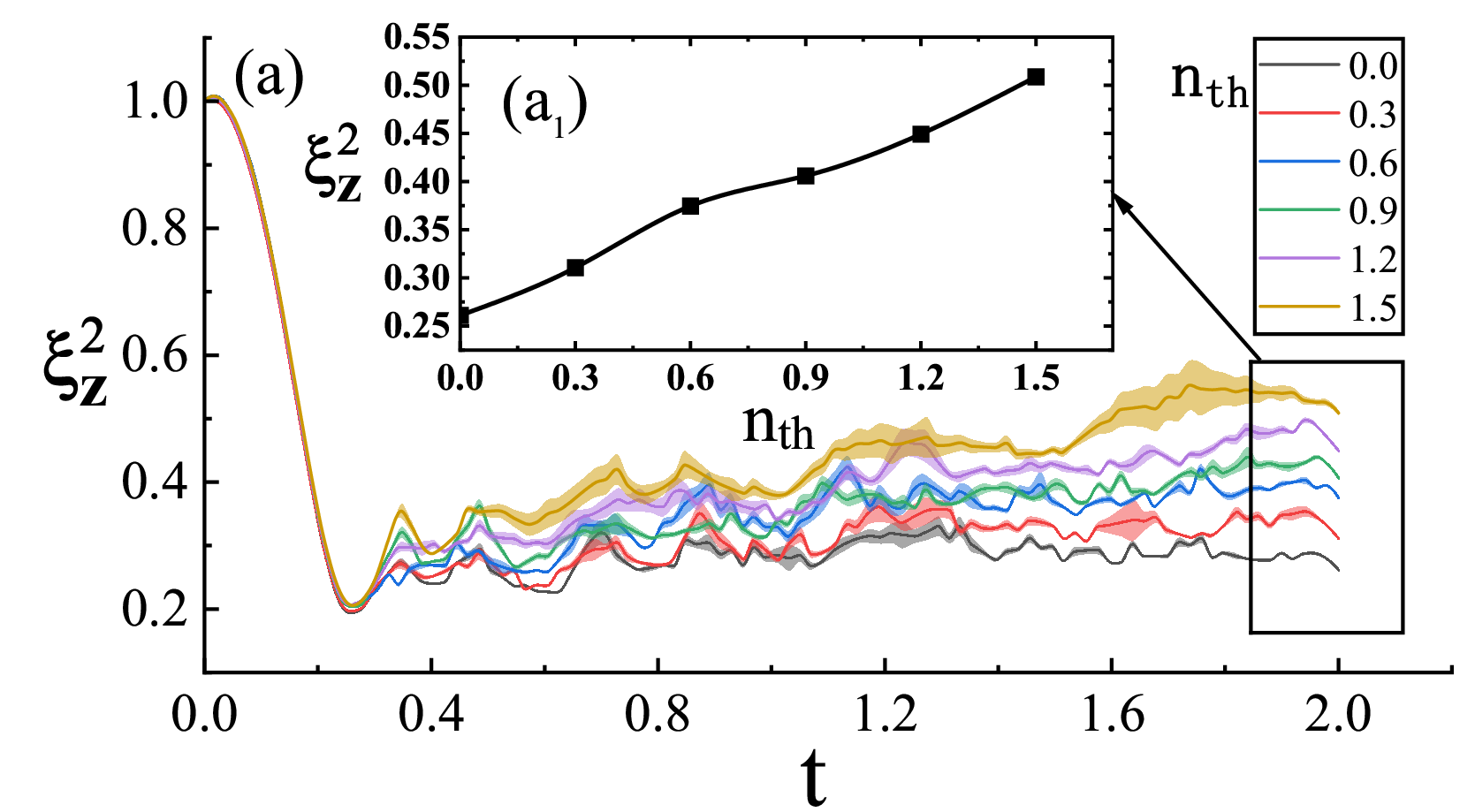}
\caption{The average evolutions of 30 squeezing parameters with
error zones for different thermal excitations in the reservoir
with $n_{th}=\frac{1}{\exp^{\hbar\omega/k_BT}-1}$: the average
number of photons for a mode with frequency $\omega$ in the
reservoir. The samples are picked in the same manner as those
shown in Fig.~\ref{DoubleCFeacc} (a). The subgraph shows the
squeezing parameter versus average thermal excitation at the
scaled time $t=2$. The other parameters are the same as those
in Fig.~\ref{DoubleCFxi} (a).}
\label{RLlearning3temp}
\end{figure}

In the previous results, the environmental temperature was assumed
to be zero. To investigate the robustness of the proposed control
scheme, it is necessary to check the influence of temperature on
the control result. Since the temperature is positively correlated
with the average number of thermal excitations in the reservoir,
denoted by $n_{th}$, it reflects the strength of the decoherence.
Fig.~\ref{RLlearning3temp} illustrates that an incremental rise of
the thermal excitation impairs the efficacy of the control strategy
from the view of spin squeezing and the control variance.

\subsection{The comparison with current protocols}

In the two-component weakly interacting Bose-Einstein
condensate~\cite{Nature40963}, the minimum value for the squeezing
parameter $\xi^2 = N (\Delta \hat{J}_{\vec n_1})^2/(\langle \hat{J}_{\vec n_2} \rangle^2 +\langle \hat{J}_{\vec n_3}\rangle^2)$ approaches $\frac{1}{2}{\left(3/N\right)}^{2/3}$ at an optimized direction.
In the work~\cite{Nature4551216}, a BEC of thousands $^{87}$Rb atoms
can be described by the one-axis twisting Hamiltonian $\hat{H} = E_C/2 \hat{J}_z^2 - 2E_J/N \hat{J}_x$, where the Josephson energy $E_J$ characterizes the tunneling
between the two condensates, and the charging energy $E_C$ describes the
repulsive interaction inside BEC. The squeezing parameter
$\xi^2_N=\xi_{R}^{2}\cos\varphi^2=\Delta \hat{J}_z^2 / \Delta \hat{J}_{z,{\rm ref}}^2$ is used to describe squeezing, where $\varphi$ is the phase difference of the
two modes. The concurrent presence of number squeezing and high phase
coherence enabled the spin squeezing parameter $\xi_R^2 = -3.8$\,dB for
the two main well pairs of a six-well lattice (approximately 2,200 atoms),
and $\xi_R^2=-2.3$\,dB for the double-well configuration
(approximately 1,600 atoms). In a similar
BEC of $^{87}$Rb (total number 2,300) nonlinear Ramsey interferometer experiments~\cite{Nature4641165}, the minimal number squeezing factor $\xi_{N}^{2}=4\left(\Delta
\hat{J}_{z}\right)^{2}/N=-8.2_{-1.2}^{+0.8}\,\text{dB}$ after taking the
technical noise due to coupling-pulse imperfections and magnetic field
fluctuations into account. In another work~\cite{Nature4641170}, spin
squeezing was generated in an atom chip by controlling the elastic
collision interactions, resulting to $\xi_{R}^{2}=-2.5_{0.6}^{-0.6}\,\text{dB}$
for 1250$\pm$45 atoms.

Theoretically, a scheme using repeated Rabi pulses transforming
one-axis-twisting model to two-axis-twisting type with
the Hamiltonian $\hat{H}(t) = \chi \hat{J}_z^2+\Omega(t)\hat{J}_y$,
and the squeezing
parameter $\xi^2 = 2(\Delta \hat{J}_{min})^2/J = 4(\Delta \hat{J}_{min})^2/N$
can approach the Heisenberg limited $1/N$ \cite{PRL107013601}. In
another protocol using the chopped random
basis technique~\cite{PRA93013851}, the squeezing parameter
$\xi^2 = N(\Delta \hat{J}_x)^2/|\langle \hat{J}_z\rangle|^2 \sim 2.1/N^{0.94}$
with a time-random-fluctuation Hamiltonian. In a
four-step strategy including constant-value and time-varying
controls to prepare spin squeezing for an ensemble of the two-level
systems coupled to a single-mode bosonic field, when $N=2J=8$,
the squeezing parameter $\xi^2 = 2(\Delta \hat{J}_{min})^2/J$ can
approach $-8$\,dB~\cite{PRA103032601}. In these proposals, the optimized
spin squeezing occurs at certain time and phase angles.

According to our results in Fig.~\ref{DoubleCFeacc}, Fig.~\ref{RLlearning3size},
and Fig.~\ref{RLlearning3temp}, the squeezing parameter $\xi_{Z}^2=N\Delta \hat{J}_{z}^2/|\langle \hat{J}_s\rangle|^2$ can be stabilized in the range
of $-5$\,dB to $-7$\,dB for a long time. Compared with the current works,
the advantage of our scheme lies in its ability to provide spin-squeezed
states along $\hat{J}_z$ over a longer time window without the need to
extract squeezed states at a specific squeezing time. Moreover, we have
considered quantum mechanical dissipation, making our theoretical research
closer to actual experimental scenarios.

\section{Discussions}\label{discussion}

\subsection{Experimental feasibility}\label{exp}
As mentioned above, atomic BEC are promising platforms
for implementing the control protocol. In the current experiments, the
condensate ensemble can be effectively modeled as the collective
pseudo-spin model described by ~(\ref{HC})~\cite{PRL99170405,Nature4641165,Nature4641170}. In these experiments, hyperfine states of the condensed atoms play the
role of the spin-up and spin-down states. The nonlinear term $\hat{J}_z^2$
is contingent upon the normalized density overlap of the two BEC
components and can be modulated via Feshbach resonance. The Rabi
frequency $\Omega (t)$ is essential in our proposal, and can be
realized by a $\pi/2$ microwave pulse coupling the near-resonant
two-photon hyperfine states of a \(^{87}\)Rb BEC trapped in optical lattice~\cite{Nature4641165,Nature4641170}. The Rabi frequency can be
turned rapidly between 0 and 2$\pi\times600$Hz, enabling swift state manipulation~\cite{Nature4641165,Nature4641170}. Our protocol requires
no additional modifications to existing experimental setups, except
for a timing sequence of the Rabi pulses. Therefore, we anticipate
that our protocol can be implemented with current experimental techniques.

\subsection{Application to continuous variable
systems}\label{ContinuousC}
Continuous variable systems, characterized by their extensive degrees
of freedom, typically present greater challenges in terms of control.
Nonetheless, it has been demonstrated that Deep DRL can effectively
address continuous-space control challenges even in the presence of
measurement back-action noise~\cite{PRL125100401}. Continuous position
measurement on the particle leads to state reduction and can be
stabilized employing measurement-based feedback control~\cite{PLA129419}.
For DRL, the continuous position measurement acts as the input of the
neural network. To design the control field, the differential equation
of the expectations would be used, instead of the quantum density matrix
directly. Although the controlled system and quantities may be continuous,
the control field can be pulsed.

\subsection{Application to other quantum system}\label{otherM}
The operations are acting on all particles identically in the
interferometer. In the large $N$ (the number of particles) limit,
the collective spin model can also be mapped to the bosonic model
by the Holstein-Primakoff transformation:
$\hat{J}_z=N/2 -\hat{c}^\dagger \hat{c}\simeq N/2$ and
$\hat{J}_+=(N-\hat{c}^\dagger \hat{c})^{1/2}\hat{c}\simeq \sqrt{N}\, \hat{c}$,
where $\hat{c}$ ($\hat{c}^\dagger$) is a bosonic annihilation (creation)
operator~\cite{PR581098}. This mapping hints that we can apply the
control proposal to such quantum systems to pursue quantum resources.

\subsection{Replaceability of the reinforcement learning module}\label{Replace}
DQN is employed as an implementation of the control scheme, actually
the RL agent can be replaced by other modules possessing similar
learning functions. Several candidates capable of substituting the
RL agent have been identified and evaluated. For example, the State-Action-Reward-State-Action algorithm~\cite{CS2500611}, Deep
Deterministic Policy Gradient~\cite{PMLR387395}, Asynchronous Advantage Actor-Critic~\cite{PMLR19281937}, Dueling Network~\cite{arXiv151106581}
and so on.

Reinforcement learning provides a tool to optimize the dynamics of
a quantum system. The optimal criterion varies for different control
targets. The principle of using this scheme is that the controlled
systems evolve under certain mapping rules (differential equations).
And the optimal module just finds the road to the optimal goal more
efficiently. After all, the machine-learning agents learn the
patterns or mappings based on the statistical distributions.

\section{Conclusion}
\label{CONC}
In conclusion, we have proposed a reinforcement-learning-based control
strategy to generate non-classical collective spin states in an open
environment. The machine learning agent can design a suite of control
sequences to prepare spin-squeezed states accompanied by entanglement.
We find larger frequency of applying the pulses enhances the performance
of the control strategy. However, more choices of actions do not
contribute obviously to the control outcomes. The scalability of this
framework is demonstrated, facilitating its applicability to larger
systems. Notably, thermal excitations progressively undermine the control
performance. The control proposal can be realized within current
experiments of atomic Bose-Einstein condensates. The extensions of the
control paradigm to continuous control problems and other quantum systems
are discussed. The versatility of this scheme is underlined by the potential
of substituting the reinforcement learning agent with other optimization
modules.

\section{Acknowledgements}
X. L. Zhao thanks discussions with Li Jiachun and Hou Shaocheng,
Natural Science Foundation of Shandong Province, China,
No.ZR2020QA078, No.ZR2023MD064, and National Natural Science
Foundation of China, No.12005110, No.12074206.

\end{document}